\documentclass[%
notitlepage,%
nofootinbib,%
amsmath,%
amssymb,%
superscriptaddress,%
12pt,
a4paper
]{revtex4-1}

\usepackage{amsmath}
\usepackage{amssymb}
\allowdisplaybreaks
\usepackage{graphicx}
\usepackage[top=25mm,bottom=30mm,left=15mm,right=15mm]{geometry}
\usepackage{cases}
\usepackage{hyperref}

\begin{document}
\renewcommand{\baselinestretch}{1.0}\normalsize

\title{Resonances in positron scattering on a supercritical nucleus
  and spontaneous production of $e^{+}e^{-}$ pairs}
\author{S.I. Godunov}
\email{sgodunov@itep.ru}
\affiliation{Alikhanov Institute for Theoretical and Experimental
  Physics, National Research Centre Kurchatov Institute, ul. Bolshaya
  Cheremushkinskaya 25, Moscow, 117218 Russia}
\affiliation{Novosibirsk State University, Novosibirsk, 630090,
  Russia}
\author{B. Machet}
\email{machet@lpthe.jussieu.fr}
\affiliation{Sorbonne Universit\'{e}s, LPTHE, Universit\'e P. et M. Curie, BP 126, 4, place Jussieu F-75252 Paris Cedex 05, France}
\affiliation{CNRS, UMR 7589, LPTHE, F-75005, Paris, France}
\author{M.I. Vysotsky}
\email{vysotsky@itep.ru}
\affiliation{Alikhanov Institute for Theoretical and Experimental
  Physics, National Research Centre Kurchatov Institute, ul. Bolshaya
  Cheremushkinskaya 25, Moscow, 117218 Russia}
\affiliation{National Research University Higher School of Economics,
  Myasnitskaya str., 20, Moscow, 101978, Russia}

\begin{abstract}
  We re-examine the physics of supercritical nuclei, specially
  focusing on the scattering phase $\delta_{\varkappa}$ and its
  dependence on the energy $\varepsilon$ of the diving electronic
  level, for which we give both exact and approximate formulas. The
  Coulomb potential $Z\alpha/r$ is rounded to the constant $Z\alpha/R$
  for $r < R$. We confirm the resonant behavior of
  $\delta_{\varkappa}$ that we investigate in details. In addition to
  solving the Dirac equation for an electron, we solve it for a
  positron, in the field of the same nucleus. This clarifies the
  interpretation of the resonances. Our results are compared with
  claims made in previous works.
\end{abstract}

\maketitle

\section{Introduction}

The Coulomb problem for a nucleus with charge $Z>Z_{\rm cr}$ was
recently analysed \cite{Kuleshov} by solving the Dirac equation for an
electron in the external field of this nucleus. Because of the
specificity of the Dirac equation that accounts simultaneously for
electrons and positrons this problem gets connected to the scattering
of positrons (holes in the Dirac sea) on the nucleus (see below). The
behavior of the scattering amplitude was found to be very peculiar: it
contains resonances and their energies, obtained from an analytical
formula found in \cite{Kuleshov},
\begin{equation}
  \varepsilon = -\xi + \frac{i}{2} \gamma, \;\;
  \xi > m, \;\;
  \gamma > 0,
  \label{eq:1}
\end{equation}
correspond to poles of the $S$ matrix located above the left cut, on
the second (unphysical) sheet of the energy plane. The resonances in
positron scattering were discussed in
Refs. \cite{MRG1:1972,MRG2:1972}.

At $Z<Z_{\rm cr}$, the width $\gamma$ vanishes, and this equation
describes the usual bound states of electrons in the Coulomb field of
the nucleus.

When $Z>Z_{\rm cr}$, $\gamma\neq0$ makes these states quasistationary
\cite{Mur:1976wh,Popov:1976dh}.

For electrons, as $Z$ increases, the transition from bound states to
resonant states corresponds to the diving of the bound states, which
start at $\varepsilon=+m$, downwards into the lower continuum.

In the present paper, in order to clarify the situation, we will also
study ``the Dirac equation for positron''. By this we mean here the
standard Dirac equation with the substitution of electron charge $e$
by $-e$.

Now, as $Z$ increases, bound states raise up from $\varepsilon=-m$ and
become resonant in the upper continuum.

For $Z<Z_{\rm cr}$, the interpretation of these bound states (also
noted in \cite{GMR} chapter 4.3) is the following. For obvious reasons
they cannot be $\left(e^{+}N^{+}\right)$ bound states, but are just
our previous $\left(e^{-}N^{+}\right)$ bound states. There is no more
information in there\footnote{The reader can be convinced as regards
  this interpretation as follows. In the present simple formalism,
  which only uses the Dirac equation, the energy of a bare positron,
  which is obtained by simply taking the limit $Z\to0$, is found to be
  $-m$. Since the ``production'' of such a particle costs at least the
  energy $+m$, the result that is obtained can only be interpreted in
  term of an electron with energy $-(-m)=+m$. This is what we mean by
  the statement that ``there is no more information''. A more
  satisfying description of positrons can only be achieved in the
  framework of Quantum Field Theory, where creating (annihilating) an
  electron and annihilating (creating) a positron both occur in the
  expansion of the field operator $\psi$ in terms of creation and
  annihilation operators.\label{footnote:pos_bound_states}}.

For $Z>Z_{\rm cr}$, we find that $\left(e^{+}N^{+}\right)$ resonances
occur at the energies
\begin{equation}
  \varepsilon_{\rm p} = \xi - \frac{i}{2} \gamma, \;\;
  \xi > m, \;\;
  \gamma > 0,
  \label{eq:2}
\end{equation}
which now correspond to poles of the $S$ matrix below the right cut of
the energy plane, also, as it should be, on the second, unphysical,
sheet. This result confirms the proposal made in \cite{Kuleshov}
that the sign of the energy in (\ref{eq:1}) should be reversed.

This change of sign we are accustomed to when dealing with holes in
the lower continuum: the absence of an electron with energy
$-\varepsilon$ is then interpreted as the presence of a positron with
energy $\varepsilon$. It is now to be operated on the empty states of
the energy levels that dive into the lower continuum. Our
consideration of the Dirac equation for positrons therefore helps to
clarify the nature and position of the resonances.

No physical interpretation for them was suggested in
\cite{Kuleshov}. It was only claimed that spontaneous $e^{+}e^{-}$
pair production by naked nuclei at $Z>Z_{\rm cr}$, as discussed in
\cite{MRG1:1972,MRG2:1972,Voronkov:1961,Gershtein:1969,Greiner:1969,Popov:1970-1,
  Popov:1970-2,Gerstein:1969-lett,Popov:1970nz,Popov:1970-ZhETF-2,
  Zeldovich:1972,Zeldovich:1971,KP:2014,Gershtein1973,Okun:1974rza,
  GMR,GMM}, does not occur.

We, however, do not see any sensible objection to the occurrence of
this process: an empty state diving into the lower continuum gets
filled by one electron of the Dirac sea; the resulting hole in the sea
is the positron that gets ejected by the nucleus the charge of which
has become $Z-1$. The characteristic time of this emission process is
$1/\gamma$, in agreement with the results obtained in
\cite{MRG1:1972,MRG2:1972,Voronkov:1961,Gershtein:1969,Greiner:1969,Popov:1970-1,
  Popov:1970-2,Gerstein:1969-lett,Popov:1970nz,Popov:1970-ZhETF-2,
  Zeldovich:1972,Zeldovich:1971,KP:2014,Gershtein1973,Okun:1974rza,
  GMR,GMM}.

Furthermore, spontaneous production of $e^+e^-$ pairs was recently
observed in the numerical solution of the Dirac equation in the case
of heavy ion collisions \cite{Maltsev:2014qna,Maltsev2017}.

The plan of the paper is as follows. In Section~\ref{sec:lower},
following \cite{Kuleshov} and using the Dirac equation, we study the
scattering of states of the lower continuum on a supercritical
nucleus. In addition to reproducing the approximate results obtained
in \cite{Kuleshov} we get explicit results without using an expansion
over the parameter $m\times R$, where $R$ is the nucleus radius. Such
an expansion being good for electrons does not work for heavy
particles, for example, muons~\cite{Mur:1976wh,Popov:1976dh}. In
Section~\ref{sec:upper}, we use instead the Dirac equation for
positrons (see above) and study the scattering of states of its upper
continuum on a supercritical nucleus. We conclude in
Section~\ref{sec:conclusions}.

\section{Lower continuum wave functions and scattering
  phases in the Coulomb field of a supercritical
  nucleus}
\label{sec:lower}

The radial functions of the Dirac equation $F(r) \equiv rf(r)$ and
$G(r) \equiv rg (r)$ are determined by the following differential
equations \cite{Bethe,Bethe2,BLP}:
\begin{equation}
  \left\{
  \begin{aligned}
    &\frac{dF}{dr} + \frac{\varkappa}{r}F -
    \left(\varepsilon + m - V(r)\right)G = 0,\\
    &\frac{dG}{dr} - \frac{\varkappa}{r}G +
    \left(\varepsilon - m - V(r)\right)F = 0,
  \end{aligned}
  \right.
  \label{eq:3}
\end{equation}
where $\varkappa = -(j+1/2) = -1, -2,\dots$ for $j = l + 1/2$ and
$\varkappa = (j +1/2)= 1,2,3\dots$ for $j = l-1/2$ and the ground
state corresponds to $\varkappa = -1$ (let us note that in
\cite{Kuleshov} the Dirac equation with the substitution
$F\Rightarrow -F$ is used).

In order to deal with the case $Z\alpha >1$ the Coulomb potential
should be regularised at $r=0$ \cite{PomSmo:1945}.  To do this we
shall approximate the nucleus as a homogeneous charged sphere with
radius $R$ (the so-called rectangular cutoff). Thus, the potential in
which the Dirac equation should be solved looks like:
\begin{subequations}
  \begin{numcases}{V(r)=}
    -\frac{Z\alpha}{R}, & $r < R$,
    \label{eq:potential_r<R}\\
    -\frac{Z\alpha}{r}, & $r > R$.
    \label{eq:potential_r>R}
  \end{numcases}
  \label{eq:potential}
\end{subequations}

At small distances $r< R$, substituting expression
(\ref{eq:potential_r<R}) into (\ref{eq:3}), we obtain the Dirac
equation with a constant potential, the solution of which is expressed
through Bessel functions. In order to obtain finite $f$ and $g$ at
$r=0$ among the two sets of solutions the one with a positive index of
the Bessel function should be selected\footnote{Any solution with a
  negative index of the Bessel function is not normalizable, so it
  should be discarded.}:
\begin{equation}
  \left(
    \begin{array}{l}
      F \\
      G
    \end{array}\right) = {\rm const}\cdot\sqrt{\beta r}\cdot
  \left(
    \begin{array}{l}
      \mp J_{\mp(1/2 + \varkappa)} (\beta r)\\
      J_{\pm(1/2 - \varkappa)} (\beta r)
      \frac{\beta}{\varepsilon+m+\frac{Z\alpha}{R}}
    \end{array}
  \right),\; r < R,
  \label{5}
\end{equation}
where $\beta = \sqrt{(\varepsilon + Z\alpha /R)^2 - m^2}$. Upper
(lower) signs should be taken for $\varkappa < 0$ ($\varkappa > 0$).

For $r>R$, we need the solution of the Dirac equation for the Coulomb
potential. We introduce the standard quantity $\lambda$ which, for
$-m < \varepsilon < m$, equals
$\lambda = \sqrt{(m-\varepsilon)(m+\varepsilon)} \equiv -ik$, where
$k$ is the electron momentum. Here we have to make an important
remark. Since later we are going to look for resonances in the complex
$\varepsilon$ plane, we must carefully define the square roots used
here. Each of them, $\sqrt{m-\varepsilon}$ and $\sqrt{m+\varepsilon}$,
are defined on two Riemann sheets of the complex $\varepsilon$
plane. To avoid ambiguous expressions let us introduce a uniquely
defined function ${\rm sqrt(z)}$ as follows:
\begin{equation}
  \label{eq:sqrt_general_0}
  {\rm
    sqrt}\left(|z|e^{i{\rm Arg}(z)}\right)=\sqrt{|z|}e^{i{\rm Arg}(z)/2},\text{
    for } {\rm Arg}(z)\in(-\pi;\pi].
\end{equation}
For example
\begin{equation}
  \label{eq:sqrt_0}
  {\rm sqrt}\left(z\right)=
  \begin{cases}
     i& \text{for }z=-1+i\cdot0,\\
     i& \text{for }z=-1,\\
    -i& \text{for }z=-1-i\cdot0.
  \end{cases}
\end{equation}
It is therefore the first branch of the function $\sqrt{z}$ with the
cut $(-\infty;0)$. The second branch is given by $-{\rm
  sqrt}(z)$. This definition is also very convenient because the
square root is defined in this way in many numerical tools for
computers.

\begin{figure}[t]
  \centering
  \includegraphics[width=4.5in]{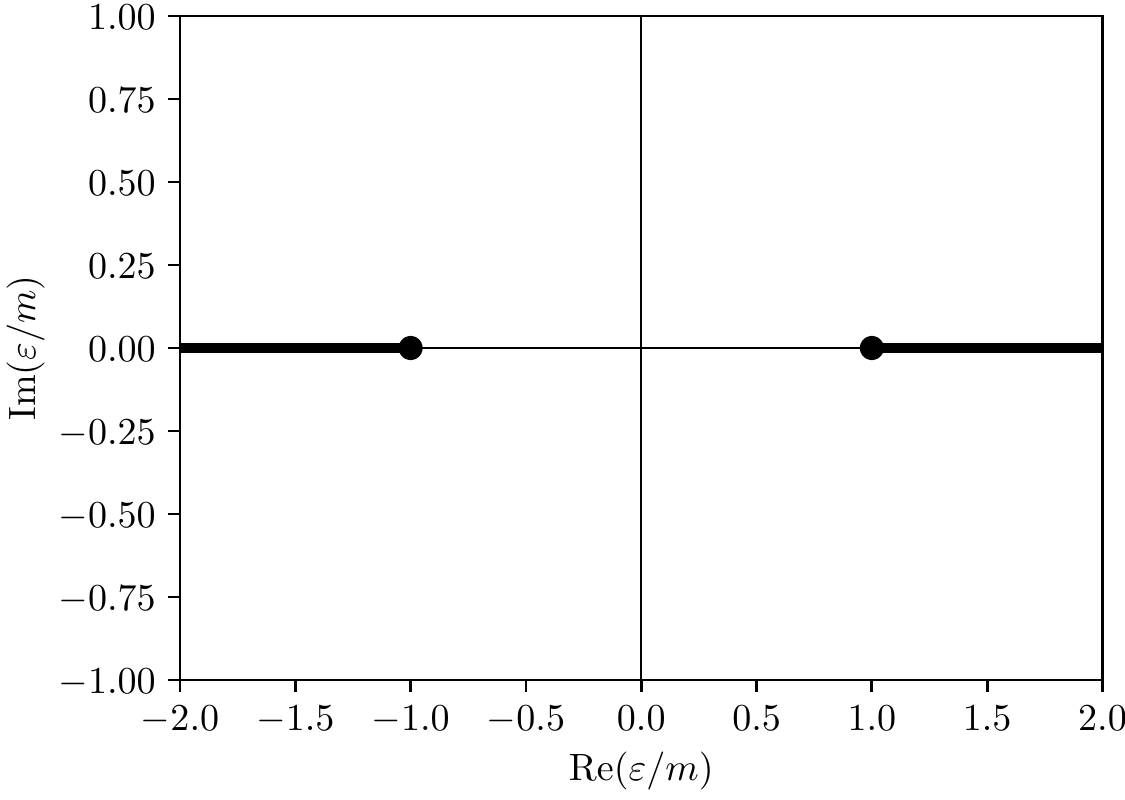}
  \caption{The plane of complex energy $\varepsilon$.}
  \label{fig:cuts}
\end{figure}

Switching branches of both square roots, $\sqrt{m-\varepsilon}$ and
$\sqrt{m+\varepsilon}$, leads to the same value of
$\lambda$. Therefore, $\lambda$ is defined on the two Riemann sheets
according to:
\begin{equation}
  \label{eq:lambda_sheets}
  \lambda=
  \begin{cases}
    {\rm sqrt}\left(m-\varepsilon\right)\cdot
    {\rm sqrt}\left(m+\varepsilon\right)&
    \text{on the physical sheet,}\\
    -{\rm sqrt}\left(m-\varepsilon\right)\cdot
    {\rm sqrt}\left(m+\varepsilon\right)&
    \text{on the unphysical sheet,}
  \end{cases}
\end{equation}
with two cuts originating, respectively, from each of the square roots
(see Fig.~\ref{fig:cuts})\footnote{The procedure used in
  \cite{Kuleshov} amounts to stating that, below the left cut,
  $\lambda = -i\sqrt{(m-\varepsilon)(-m-\varepsilon)}$. So doing,
  $\sqrt{-m-\varepsilon}$ is defined with the same cut $(-\infty;-m)$
  as $\sqrt{m+\varepsilon}$, with positive values below the cut. With
  such a definition,
  $-i\sqrt{-m-\varepsilon}={\rm sqrt}\left(m+\varepsilon\right)$
  everywhere on the physical sheet, not only below the left cut. There
  is no need to rewrite formulas in this way since, when numerical
  outputs are needed, we should return to the original
  definition~(\ref{eq:lambda_sheets}). Let us note that on the first
  sheet formulas (17) and (26) from \cite{Kuleshov} are exactly the
  same.}. From general arguments of scattering theory, we know that
electron bound states are located at real $\varepsilon$ in the
interval $-m < \varepsilon < m$.  Unbound electron states are located
above the right cut and unbound positron states below the left cut.

In what follows we shall use the following conventions for the
``$\sqrt{~}$'' symbol:
\begin{align}
  \label{eq:roots}
  \sqrt{m+\varepsilon}&=
  \begin{cases}
    {\rm sqrt}\left(m+\varepsilon\right)&\text{on the physical sheet,}\\
    -{\rm sqrt}\left(m+\varepsilon\right)&\text{on the unphysical sheet,}
  \end{cases}\\
  \sqrt{m-\varepsilon}&={\rm sqrt}\left(m-\varepsilon\right)\text{ on both sheets.}
\end{align}
It does not matter which root changes sign when we go to the second
sheet since we can always also change the signs of both.

We are looking for solution written in the standard form
\cite{BLP}\footnote{Let us note that changing the signs of both square
  roots is still permitted since it leads to changing the sign of the
  full wave function.}:
\begin{equation}
  \left(
    \begin{array}{c}
      F \\
      G
    \end{array}
  \right) = \left(
    \begin{array}{c}
       \sqrt{m+\varepsilon} \\
      -\sqrt{m-\varepsilon}
    \end{array}
  \right) {\rm exp} (-\rho/2)\rho^{i\tau} \left(
    \begin{array}{c}
      Q_1 + Q_2 \\
      Q_1 - Q_2
    \end{array}
  \right),
  \label{8}
\end{equation}
where $\tau=\sqrt{\left(Z\alpha\right)^{2}-\varkappa^{2}}$,
$\rho = 2\lambda r = -2ikr$, $Q_1$ and $Q_2$ are determined by
differential equations, the solutions of which are Kummer confluent
hypergeometric functions $_1F_1(\alpha,\beta,z)$ (also sometimes noted
$F(\alpha,\gamma,z)$ like in \cite{BLP}). In textbooks dealing with
the case $Z \alpha <1$, $R=0$, only solutions regular at $r=0$ are
considered. We must instead here take into account both type of
solutions of the equations for $Q_1$ and $Q_2$. The formulas for the
$Q_i$ are derived in Appendix~\ref{sec:Q1Q2}. From (\ref{A6}) to
(\ref{A8}) we get:
\begin{equation}
  \hspace{-7mm}\left\{
    \begin{aligned}
      Q_1 &= 
      C\cdot\frac{-\frac{iZ\alpha m}{k}+\varkappa}
      {-i\tau+\frac{iZ\alpha\varepsilon}{k}}
      \cdot {_1F_{1}}\left(i\tau-\frac{iZ\alpha\varepsilon}{k},2i\tau +1,\rho\right)+\\ 
      &\hspace{40mm}+D\cdot\frac{-\frac{iZ\alpha m}{k} +\varkappa}
      {i\tau+\frac{iZ\alpha\varepsilon}{k}} \rho^{-2i\tau}{_{1}F_{1}}\left(-i\tau - 
        \frac{iZ\alpha\varepsilon}{k}, -2i\tau+1, \rho\right),\\
      Q_2 &= C\cdot{_{1}F_{1}}\left(1+i\tau - \frac{iZ\alpha\varepsilon}{k},
        2i\tau+1, \rho\right) + D\rho^{-2i\tau}{_{1}F_{1}}\left(1-i\tau -
        \frac{iZ\alpha\varepsilon}{k}, -2i\tau+1, \rho\right),
    \end{aligned}\right.\label{eq:9}
\end{equation}
where $C$ and $D$ are arbitrary coefficients\footnote{Unlike in
  \cite{Kuleshov} we did not feel necessary to use Tricomi
  functions.}.

The scattering phase $\delta_{\varkappa}(\varepsilon,Z)$ is determined
by investigating the behavior of the wave function at large $r$. To this
purpose, the asymptotic expansion of $_1F_1$ at large $|z|$
\begin{equation}
  _{1}F_{1}(\alpha, \gamma, z)\Big\rvert_{|z| \to \infty} =
  \frac{\Gamma(\gamma)}{\Gamma(\gamma-\alpha)}(-z)^{-\alpha} 
  [1+ O(1/z)]
  + \frac{\Gamma(\gamma)}{\Gamma(\alpha)} e^z z^{\alpha -\gamma} [1+O(1/z)]
  \label{11}
\end{equation}
is very useful.

Using the asymptotic expansion (\ref{11}) for the Kummer functions
occurring in (\ref{eq:9}) gives:
\begin{align}
  \left.\left(
    \begin{array}{c}
      F \\
      G
    \end{array} 
  \right)\right|_{r\to\infty} &= A\cdot\left(
    \begin{array}{c}
       \sqrt{m+\varepsilon} \\
      -\sqrt{m-\varepsilon}
    \end{array}
  \right)\times\\
  \times\Biggl(C&\left[
    e^{-\frac{\rho}{2}}\frac{\Gamma\left(2i\tau+1\right)}
    {\Gamma\left(1+i\tau+\frac{iZ\alpha\varepsilon}{k}\right)}
    \frac{\frac{iZ\alpha m}{k}-\varkappa}{i\tau-\frac{iZ\alpha\varepsilon}{k}}
    \rho^{i\tau}\left(-\rho\right)^{-i\tau}
    \left(-\rho\right)^{\frac{iZ\alpha\varepsilon}{k}}
    \pm
    e^{\frac{\rho}{2}}\frac{\Gamma\left(2i\tau+1\right)}
    {\Gamma\left(1+i\tau-\frac{iZ\alpha\varepsilon}{k}\right)}
    \rho^{-\frac{iZ\alpha\varepsilon}{k}}
    \right]
    +\\
  +D&\left[
    e^{-\frac{\rho}{2}}\frac{\Gamma\left(-2i\tau+1\right)}
    {\Gamma\left(1-i\tau+\frac{iZ\alpha\varepsilon}{k}\right)}
    \frac{\frac{iZ\alpha m}{k}-\varkappa}{-i\tau-\frac{iZ\alpha\varepsilon}{k}}
    \rho^{-i\tau}\left(-\rho\right)^{i\tau}
    \left(-\rho\right)^{\frac{iZ\alpha\varepsilon}{k}}
    \pm
    e^{\frac{\rho}{2}}\frac{\Gamma\left(-2i\tau+1\right)}
    {\Gamma\left(1-i\tau-\frac{iZ\alpha\varepsilon}{k}\right)}
    \rho^{-\frac{iZ\alpha\varepsilon}{k}}
    \right]\Biggr),
  \label{12}
\end{align}
where the upper sign corresponds to $F$ and the lower sign corresponds
to $G$.

The ratio
\begin{align}
  \frac{\left(-\rho\right)^{\frac{iZ\alpha\varepsilon}{k}}}
  {\rho^{-\frac{iZ\alpha\varepsilon}{k}}}
\end{align}
yields the Coulomb logarithm (for real $\varepsilon$ below the left
cut it gives
$\exp\left[\frac{2iZ\alpha\varepsilon}{k}\ln\left(2kr\right)\right]$). Since
the latter does not contribute to the differential scattering cross
section at nonzero angle $\theta$, we will omit this term in our
further calculations.

From the general formula
\begin{align}
  \left.\left(
  \begin{array}{c}
    F \\
    G
  \end{array} 
  \right)\right|_{r\to\infty} \propto\left(
  \begin{array}{c}
    \sqrt{m+\varepsilon} \\
    -\sqrt{m-\varepsilon}
  \end{array}
  \right)
  \left\{e^{i(kr+\frac{Z\alpha\varepsilon}{k}\ln(2kr))} e^{2i\delta}
  \pm e^{-i(kr+\frac{Z\alpha\varepsilon}{k}\ln(2kr))}\right\}
\end{align}
it follows that the ratio of the remaining coefficients define the
scattering phase $\delta_{\varkappa}$ (on the real axis below the left
cut $e^{-\rho/2}\equiv e^{ikr}$ corresponds to the outgoing wave and
$e^{\rho/2}\equiv e^{-ikr}$ corresponds to the incoming wave):
\begin{align}
  \label{eq:phase}
  e^{2i\delta_{\varkappa}}
  &=-\frac{1}{\varkappa+\frac{iZ\alpha m}{k}}\cdot
    \frac
    {\frac{C}{D}\cdot\frac{\Gamma\left(2i\tau\right)}
    {\Gamma\left(i\tau+\frac{iZ\alpha\varepsilon}{k}\right)}
    \rho^{i\tau}\left(-\rho\right)^{-i\tau}
    -
    \frac{\Gamma\left(-2i\tau\right)}
    {\Gamma\left(-i\tau+\frac{iZ\alpha\varepsilon}{k}\right)}
    \rho^{-i\tau}\left(-\rho\right)^{i\tau}}
    {\frac{C}{D}\cdot\frac{\Gamma\left(2i\tau\right)}
    {\Gamma\left(1+i\tau-\frac{iZ\alpha\varepsilon}{k}\right)}
    -
    \frac{\Gamma\left(-2i\tau\right)}
    {\Gamma\left(1-i\tau-\frac{iZ\alpha\varepsilon}{k}\right)}},
\end{align}
where
\begin{align}
  \rho^{i\tau}\left(-\rho\right)^{-i\tau}
  &=
    \exp\left[i\tau\ln\left(\rho\right)-i\tau\ln\left(-\rho\right)\right]
    =\exp\left[-\tau\left(
    {\rm Arg}\left[\rho\right]-{\rm Arg}\left[-\rho\right]
    \right)\right]=\\
  &=e^{-\pi\tau\cdot{\rm sign}\left[{\rm Arg}\left[\rho\right]\right]}.\nonumber
\end{align}

The resonance of the scattering amplitude corresponds to the pole of
the $S$-matrix element $S\equiv e^{2i\delta}$ and from
(\ref{eq:phase}) we immediately get an equation for the position of
this pole in the $\varepsilon$-plane:
\begin{align}
  \label{eq:pole}
  \frac{C}{D}\cdot\frac{\Gamma\left(2i\tau\right)}
  {\Gamma\left(1+i\tau-\frac{iZ\alpha\varepsilon}{k}\right)}
  -
  \frac{\Gamma\left(-2i\tau\right)}
  {\Gamma\left(1-i\tau-\frac{iZ\alpha\varepsilon}{k}\right)}=0.
\end{align}

In what follows we will match the solutions at $r<R$ and $r>R$ to
obtain the ratio $C/D$, such that we can calculate the phase
$\delta_{\varkappa}$ and find the poles of the $S$ matrix which
correspond to the energy levels. This procedure can be performed both
exactly and approximately.

\subsection{Exact results}

With the help of the exact formulas (\ref{5}), (\ref{8}), and
(\ref{eq:9}) we get the ratio $C/D$ from matching $F/G$ at $r=R+0$ and
$r=R-0$:
\begin{equation}
  \label{eq:CD_exact}
  \frac{C}{D}=
  -\rho_{0}^{-2i\tau}\cdot\frac{F_{g}^{-}-MF_{f}^{-}}{F_{g}^{+}-MF_{f}^{+}},
\end{equation}
where
\begin{align}
  M&=\pm\frac{\sqrt{m+\varepsilon}}{\sqrt{m-\varepsilon}}
     \cdot
     \frac{J_{\pm\left(1/2-\varkappa\right)}\left(\beta R\right)}
     {J_{\mp\left(1/2+\varkappa\right)}\left(\beta R\right)}
     \cdot
     \frac{\beta}{\varepsilon+m+\frac{Z\alpha}{R}},\label{eq:M}\\
  \label{eq:Ff}
  F_{f}^{\pm}&=
  {_{1}F_{1}}\left(\alpha_{1}^{\pm},\gamma^{\pm},\rho_{0}\right)
  \frac{\frac{iZ\alpha m}{k}-\varkappa}{\alpha_{1}^{\pm}}
  +{_{1}F_{1}}\left(\alpha_{2}^{\pm},\gamma^{\pm},\rho_{0}\right),\\
  \label{eq:Fg}
  F_{g}^{\pm}&=
  {_{1}F_{1}}\left(\alpha_{1}^{\pm},\gamma^{\pm},\rho_{0}\right)
  \frac{\frac{iZ\alpha m}{k}-\varkappa}{\alpha_{1}^{\pm}}
  -{_{1}F_{1}}\left(\alpha_{2}^{\pm},\gamma^{\pm},\rho_{0}\right),
\end{align}
and
\begin{equation}
  \label{eq:args}
  \alpha_{1}^{\pm}=\pm i\tau-\frac{iZ\alpha\varepsilon}{k},
  \alpha_{2}^{\pm}=1\pm i\tau-\frac{iZ\alpha\varepsilon}{k},
  \gamma^{\pm}=\pm 2i\tau+1,
  \rho_{0}=-2ikR.
\end{equation}
The numerical evaluation of the square roots in (\ref{eq:M}) and of
$k$ in (\ref{eq:args}) for \emph{real} $\varepsilon$ is somewhat
tricky since one should carefully choose the side of the cut to
use. Due to the definition
(\ref{eq:sqrt_general_0})--(\ref{eq:sqrt_0}) of the ${\rm sqrt}()$
function the expression ${\rm sqrt}\left(m+\varepsilon\right)$ gives,
for \emph{real} $\varepsilon$, the values above the cut such that
$-{\rm sqrt}\left(m+\varepsilon\right)$ should be used. It corresponds
\emph{formally} to calculating the scattering phase on the second
(unphysical) sheet. The same holds for $k$. For any \emph{real}
$\varepsilon$ it is also possible to use
$k={\rm sqrt}\left(\varepsilon^{2}-m^{2}\right)$ which chooses the
correct side of the cut; then
$\sqrt{m+\varepsilon}/\sqrt{m-\varepsilon}=-ik/\left(m-\varepsilon\right)$.

With the help of (\ref{eq:CD_exact}) we can calculate the scattering
phase $\delta_{\varkappa}$ defined by (\ref{eq:phase}).

In the domain $\varepsilon<-m$, $\delta_{\varkappa}(\varepsilon,Z)$
gives the scattering phase of a positron with energy
$\varepsilon_{\rm p}=-\varepsilon>m$ on the nucleus (for real
$\varepsilon<-m$ we get ${\rm Arg}\left[\rho\right]<0$). Its
dependence on $\varepsilon_{\rm p}$ for $\varkappa = -1$ and $Z = 232$
is shown in Fig.~\ref{fig:phase} (compare with Fig.~3 of
\cite{Kuleshov}). The scattering phase $\delta_{\varkappa}$ exhibits a
resonance behavior; it goes through $\pi/2$ at
$\varepsilon_{\rm p}/m\approx 5.06$.

We obtain the equation for the position of the poles by substituting
(\ref{eq:CD_exact}) into (\ref{eq:pole})
\begin{align}
  \label{eq:pole_detailed}
  \frac{\Gamma\left(-2i\tau\right)}
  {\Gamma\left(2i\tau\right)}\cdot
  \frac{\Gamma\left(1+i\tau-\frac{iZ\alpha\varepsilon}{k}\right)}
  {\Gamma\left(1-i\tau-\frac{iZ\alpha\varepsilon}{k}\right)}=
  -\rho_{0}^{-2i\tau}\cdot\frac{F_{g}^{-}-MF_{f}^{-}}{F_{g}^{+}-MF_{f}^{+}}.
\end{align}
The solutions of (\ref{eq:pole_detailed}) can be found by scanning the
complex $\varepsilon$ plane.  This is the method that we used to find
the exact positions\footnote{Due to the unitarity of the $S$ matrix,
  there is a zero of $e^{2i\delta_{\varkappa}}$ at
  $\varepsilon=-\xi-\frac{i}{2}\gamma$ that is symmetric to the pole
  $\varepsilon=-\xi+\frac{i}{2}\gamma$ with respect to the real
  axis. It corresponds to incoming waves instead of the outgoing waves
  that we selected.} of the $S$-matrix poles (see
Fig.~\ref{fig:e_of_Z} and Table~\ref{tab:e_of_Z}). The energies
$\varepsilon$ of the quasistationary states are located above the left
cut on the second sheet of the complex $\varepsilon$ plane\footnote{It
  corresponds to ${\rm Re}[k]>0$,
  ${\rm Im}[k]=\frac{\displaystyle{\rm Re}[\varepsilon]{\rm
      Im}[\varepsilon]} {\displaystyle{\rm Re}[k]}<0$. In
  \cite{MRG1:1972,MRG2:1972,Zeldovich:1971} the resonance in positron
  scattering on a supercritical nucleus was discussed.}:
\begin{equation}
  \varepsilon = -\xi + \frac{i}{2}\gamma,~\xi>m,~\gamma>0.
\label{eq:resonances}
\end{equation}

\subsection{Approximate results}

In \cite{Kuleshov} the approximation $1/R\gg\varepsilon,m$ was
used. In this section we are going to reproduce their results and
compare them to the exact ones.

Being interested in the case $Z\alpha \gtrsim 1$ and taking into
account the smallness of the nucleus radius in comparison with the
electron Compton wavelength $1/m$ we obtain that
$\beta \approx Z\alpha/R$ in (\ref{5}).

The solution of the system (\ref{eq:3}) at $r > R$ should match
(\ref{5}) at $r=R$, in particular the ratio $F/G$ of both solutions at
$r=R$ should coincide. Substituting (\ref{eq:potential_r>R}) in
(\ref{eq:3}) at $r\to 0$ we easily get
\begin{equation}
  \left.\left(
      \begin{array}{l}
        F \\
        G
      \end{array}\right)\right|_{r \to 0}
  = \eta_\sigma r^\sigma \left(
    \begin{array}{c}
      -1 \\ 
      \frac{Z\alpha}{\sigma - \varkappa}
    \end{array}
  \right) + 
  \eta_{-\sigma} r^{-\sigma} \left(
    \begin{array}{c}
      -1 \\ 
      \frac{Z\alpha}{-\sigma - \varkappa}
    \end{array}
  \right),
  \label{6}
\end{equation}
where $\sigma = \sqrt{\varkappa^2 - Z^2\alpha^2}$ and $\eta_\sigma$
and $\eta_{-\sigma}$ are arbitrary constants.
Matching the ratios $F/G$ from (\ref{6}) and (\ref{5}) at $r=R$ we
obtain
\begin{equation}
  \frac{\eta_\sigma}{\eta_{-\sigma}}
  = \frac{\sigma -\varkappa}{\sigma +\varkappa}\cdot
  \frac{Z\alpha J_{\mp(1/2 +\varkappa)}(Z\alpha)
    \pm (\sigma + \varkappa)J_{\pm(1/2 -\varkappa)}(Z\alpha)}
  {Z\alpha J_{\mp(1/2 +\varkappa)}(Z\alpha)
    \mp (\sigma - \varkappa) J_{\pm(1/2 -\varkappa)}(Z\alpha)}
  \cdot\frac{R^{-\sigma}}{R^\sigma}=\tan\theta,
  \label{7}
\end{equation}
which coincides with Eq.~(13) from \cite{Kuleshov}. In the case
$Z\alpha > |\varkappa|$ one should substitute $\sigma$ by $i\tau$
(where, as before, $\tau = \sqrt{Z^2\alpha^2 - \varkappa^2}$):
\begin{equation}
  \frac{\eta_\tau}{\eta_{-\tau}}
  = \frac{i\tau -\varkappa}{i\tau +\varkappa}\cdot
  \frac{Z\alpha J_{\mp(1/2 +\varkappa)}(Z\alpha)
    \pm (i\tau + \varkappa) J_{\pm(1/2 -\varkappa)}(Z\alpha)}
  {Z\alpha J_{\mp(1/2 +\varkappa)}(Z\alpha)
    \mp (i\tau - \varkappa) J_{\pm(1/2 -\varkappa)}(Z\alpha)}
  \cdot\frac{R^{-i\tau}}{R^{i\tau}}= e^{2i\theta}.
  \label{77}
\end{equation}
The modulus of the r.h.s of (\ref{77}) can be easily checked to be
unity, this is why we can rewrite it as an
$\exp{\left(2i\theta\right)}$ with real $\theta$.

The expansion of (\ref{8}) at small
$\rho$ contains terms $\sim\rho^{i\tau}$ and
$\rho^{-i\tau}$. Comparing this expansion with (\ref{6}) and
substituting $\sigma \to i\tau$, $\eta_\sigma \to \eta_\tau$,
$\eta_{-\sigma}\to \eta_{-\tau}$ yields:
\begin{equation}
  \eta_\tau = C\cdot(-2ik)^{i\tau}
  \frac{i\tau - \varkappa + Z\alpha\sqrt{\frac{m-\varepsilon}
      {m+\varepsilon}}}{i\tau - \frac{iZ\alpha\varepsilon}{k}}, \;
  \eta_{-\tau} = D\cdot(-2ik)^{-i\tau}
  \frac{-i\tau - \varkappa + Z\alpha\sqrt{\frac{m-\varepsilon}
      {m+\varepsilon}}}{-i\tau - \frac{iZ\alpha \varepsilon}{k}}.
  \label{10}
\end{equation}

Getting an equation for $C/D$ needs matching (\ref{77}) with
$\eta_{\tau}/\eta_{-\tau}$ obtained from (\ref{10}):
\begin{align}
  \label{eq:CD}
  \frac{C}{D}=e^{2i\theta}\cdot
  \frac{\left(-2ik\right)^{-i\tau}}{\left(-2ik\right)^{i\tau}}\cdot
  \frac{Z\alpha\sqrt{m-\varepsilon}+\left(-i\tau-\varkappa\right)\sqrt{m+\varepsilon}}
  {Z\alpha\sqrt{m-\varepsilon}+\left(i\tau-\varkappa\right)\sqrt{m+\varepsilon}}\cdot
  \frac{i\tau-\frac{iZ\alpha\varepsilon}{k}}{-i\tau-\frac{iZ\alpha\varepsilon}{k}}.
\end{align}

Two sets of approximations were made in deriving (\ref{eq:CD}): i. to
get $\eta_\tau/\eta_{-\tau}$ at $r=R-0$ we replaced $\beta R$ with
$Z \alpha$ and used (\ref{6}) which was itself derived for
$Z \alpha /r \gg \varepsilon,m$; ii. to get $\eta_\tau/\eta_{-\tau}$
at $r=R+0$ we expanded (\ref{8}) and (\ref{eq:9}) at $\rho\ll 1$. For
$m\cdot R = 0.031$ one cannot expect an accuracy better than 3\% and,
with growing $|\varepsilon|$ it can even get worse. The accuracy of
the final result is not easy to guess from the start, and the best way
is to compare it with the exact solution which was found in the
previous subsection.  Note that all results in \cite{Kuleshov} are
based on the asymptotic behavior (\ref{6}) and are therefore
approximate by default.

\begin{figure}[t]
  \begin{center}
    \includegraphics[width=6.5in]{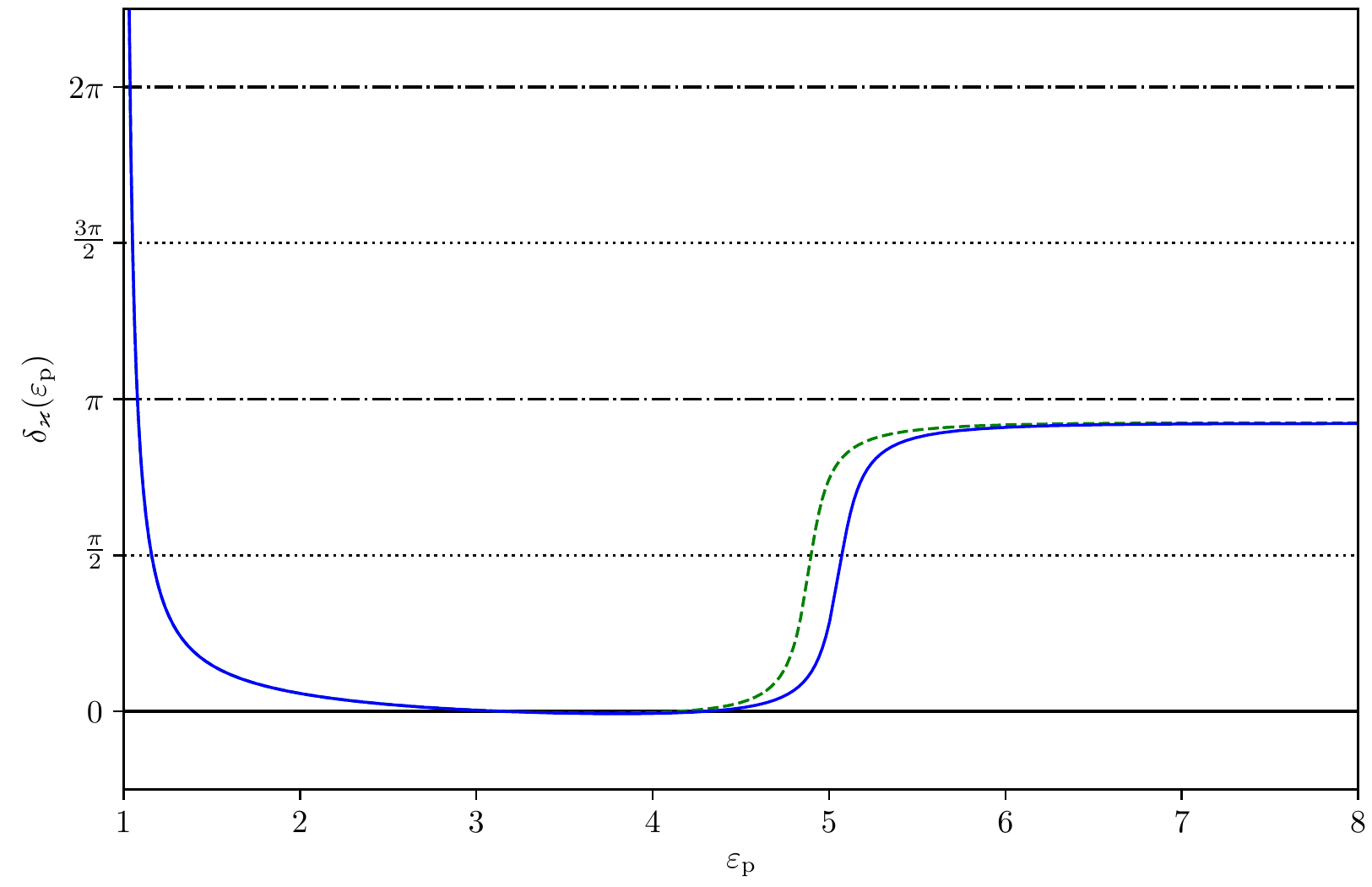}
  \end{center}
  \vspace{-10mm}
  \caption{Dependence on $\varepsilon_{\rm p}$ of the scattering phase
    $\delta_{-1}(\varepsilon_{\rm p}, 232)$ ($Z=232$ and
    $\varkappa=-1$) for a nucleus with radius $R=0.031/m$. The blue
    solid line corresponds to the exact phase, the green dashed line
    corresponds to the approximate one.}
  \label{fig:phase}
\end{figure}

Substituting (\ref{eq:CD}) into (\ref{eq:phase}) we obtain the
approximate expression for the scattering phase
$\delta_{\varkappa}$. Its dependence on
$\varepsilon_{\rm p}\equiv-\varepsilon$ for $\varkappa = -1$ and
$Z = 232$ is shown in Fig.~\ref{fig:phase} (compare with Fig.~3 of
\cite{Kuleshov}). The scattering phase $\delta_{\varkappa}$ exhibits a
resonance behavior; it goes through $\pi/2$ at
$\varepsilon_{\rm p}/m\approx 4.88$.

Let us note that on the real axis of $\varepsilon$ the expression for
the scattering phase $\delta_{\varkappa}$ can be written in the same
form as in \cite{Kuleshov} (see Appendix~\ref{sec:real_phase}).

The positions of the $S$ matrix poles are defined by the same equality
(\ref{eq:pole}) with $C/D$ given by (\ref{eq:CD}):
\begin{align}
  e^{2i\theta}=
    \frac{\left(-2ik\right)^{i\tau}}{\left(-2ik\right)^{-i\tau}}\cdot
    \frac{\Gamma\left(-2i\tau\right)}{\Gamma\left(2i\tau\right)}\cdot
    \frac{\Gamma\left(1+i\tau-\frac{iZ\alpha\varepsilon}{k}\right)}
    {\Gamma\left(1-i\tau-\frac{iZ\alpha\varepsilon}{k}\right)}\cdot
    \frac{Z\alpha\sqrt{m-\varepsilon}+\left(-i\tau+\varkappa\right)\sqrt{m+\varepsilon}}
    {Z\alpha\sqrt{m-\varepsilon}+\left(i\tau+\varkappa\right)\sqrt{m+\varepsilon}},
  \label{eq:eq_poles}
\end{align}
where the l.h.s. is defined by (\ref{77}). The r.h.s. coincides with
Eq. (26) from \cite{Kuleshov}. The exact expression
(\ref{eq:pole_detailed}) is, of course, more complicated, but, anyhow,
special functions have to be evaluated numerically in both cases.

\begin{figure}[t]
  \begin{center}
    \includegraphics[width=6.5in]{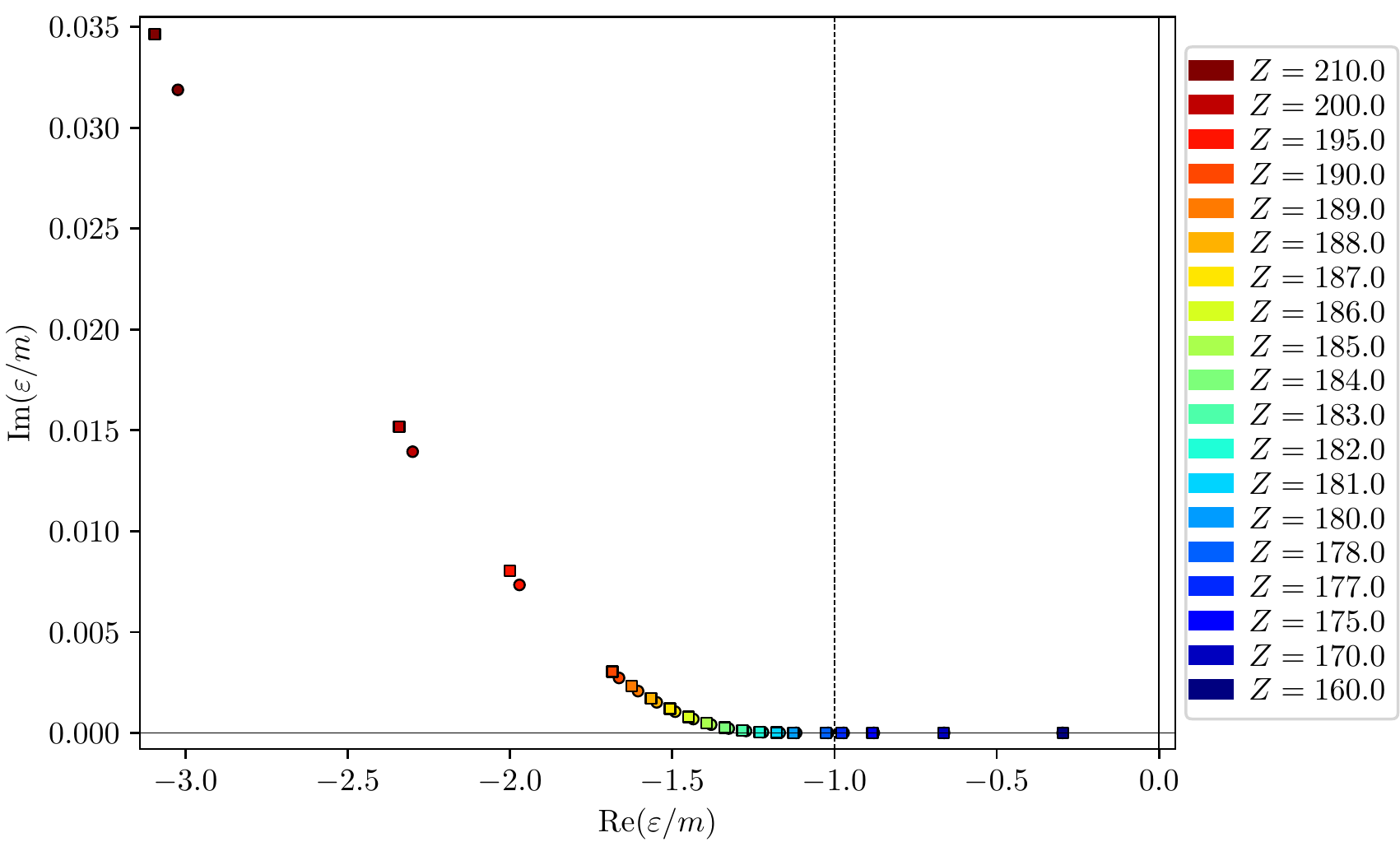}
  \end{center}
  \caption{The dependence of the ground state energy on $Z$. The
    square markers are for the exact values of the energy (see
    (\ref{eq:pole_detailed})) and the round markers are for the
    approximate ones calculated with the help of
    (\ref{eq:eq_poles}). The correspondence between color and $Z$ is
    shown in the legend (the real part of the energy is monotonically
    decreasing). At $Z=Z_{\rm cr}$ the bound states become resonances
    with positive ${\rm Im}[\varepsilon]$.}
  \label{fig:e_of_Z}
\end{figure}

The accuracy of ${\rm Re}[\varepsilon]$ obtained by the approximate
procedure is quite reasonable (see Fig.~\ref{fig:e_of_Z} and
Table~\ref{tab:e_of_Z}); however it is much worse for
${\rm Im}[\varepsilon]$, for example $\approx15\%$ at $Z=186$. This is
why it is worth getting the exact values of the energies
$\varepsilon$.

The question that we want to address now is the origin of the
resonance and how it transforms for $Z < Z_{\rm cr}$.

At $Z < Z_{\rm cr}$ the resonances become bound states, the energies
of which are determined by the same type of matching at $r=R$ as
before (it is convenient to replace now, in (\ref{eq:eq_poles}), $k$
by $i\lambda$, since on the real axis, for $-m<\varepsilon<+m$,
$\lambda$ is real positive):
\begin{equation}
  {\rm exp}(2i\theta) =
  \frac{(2\lambda)^{i\tau}}{(2\lambda)^{-i\tau}}\cdot
  \frac{\Gamma(-2i\tau)}{\Gamma(2i\tau)}\cdot 
  \frac{\Gamma\left(1+i\tau-\frac{Z\alpha\varepsilon}{\lambda}\right)}
  {\Gamma\left(1-i\tau-\frac{Z\alpha\varepsilon}{\lambda}\right)}\cdot
  \frac{Z\alpha\sqrt{m-\varepsilon}+(\varkappa-i\tau)\sqrt{m+\varepsilon}}
  {Z\alpha\sqrt{m-\varepsilon}+(\varkappa+i\tau)\sqrt{m+\varepsilon}}.
  \label{171}
\end{equation}

Last, for $Z\alpha < |\varkappa|$ we must change $i\tau$ into
$\sigma = \sqrt{\varkappa^2 - Z^2\alpha^2}$:
\begin{equation}
  \tan\theta =
  \frac{(2\lambda)^\sigma}{(2\lambda)^{-\sigma}}\cdot
  \frac{\Gamma(-2\sigma)}{\Gamma(2\sigma)}\cdot
  \frac{\Gamma\left(1+\sigma-\frac{Z\alpha\varepsilon}{\lambda}\right)}
  {\Gamma\left(1-\sigma-\frac{Z\alpha\varepsilon}{\lambda}\right)}\cdot
  \frac{Z\alpha\sqrt{m-\varepsilon}+(\varkappa-\sigma)\sqrt{m+\varepsilon}}
  {Z\alpha\sqrt{m-\varepsilon}+(\varkappa +\sigma)\sqrt{m+\varepsilon}}.
  \label{181}
\end{equation}

\begin{table}[ht!]
  \centering
  \caption{The dependence of the ground state energy on $Z$ for 
    $m\cdot R= 0.031$. The ground energy level enters the lower 
    continuum at $Z=178$ (though the imaginary part of $\varepsilon$
    is much smaller than the accuracy of the calculation). We see
    that for $Z=232$ the accuracy of the approximate formula is
    about 10\% for the ${\rm Im}[\varepsilon]$.}
  \label{tab:e_of_Z}
  \begin{tabular}[t]{|c|c|c|c|c|}
    \hline
    $Z$ &
    ${\rm Re}\left(\varepsilon_{\rm appr}\right)$ &
    ${\rm Im}\left(\varepsilon_{\rm appr}\right)$ &
    ${\rm Re}\left(\varepsilon\right)$ & 
    ${\rm Im}\left(\varepsilon\right)$ \\
    \hline
    160 & -0.297 & 0 & -0.296 & 0 \\
170 & -0.662 & 0 & -0.664 & 0 \\
175 & -0.879 & 0 & -0.883 & 0 \\
177 & -0.972 & 0 & -0.978 & 0 \\
178 & -1.020 & 0 & -1.026 & 0 \\
180 & -1.118 & 5.375e-07 & -1.127 & 9.229e-07 \\
181 & -1.169 & 6.644e-06 & -1.178 & 9.475e-06 \\
182 & -1.220 & 3.198e-05 & -1.231 & 4.168e-05 \\
183 & -1.273 & 9.562e-05 & -1.284 & 1.183e-04 \\
184 & -1.326 & 2.164e-04 & -1.338 & 2.591e-04 \\
185 & -1.380 & 4.097e-04 & -1.394 & 4.794e-04 \\
186 & -1.435 & 6.863e-04 & -1.450 & 7.903e-04 \\
187 & -1.491 & 1.053e-03 & -1.507 & 1.198e-03 \\
188 & -1.548 & 1.515e-03 & -1.565 & 1.707e-03 \\
189 & -1.605 & 2.071e-03 & -1.625 & 2.318e-03 \\
190 & -1.664 & 2.723e-03 & -1.685 & 3.030e-03 \\
195 & -1.971 & 7.335e-03 & -2.001 & 8.031e-03 \\
200 & -2.300 & 1.394e-02 & -2.341 & 1.517e-02 \\
210 & -3.023 & 3.188e-02 & -3.094 & 3.464e-02 \\
232 & -4.885 & 8.773e-02 & -5.057 & 9.638e-02 \\

    \hline
  \end{tabular}
\end{table}

Let us consider for example $Z\alpha<1$, for which taking a point-like
nucleus is reliable. At the limit $R\to0$, the r.h.s. of (\ref{7})
becomes infinite. Therefore, the spectrum of the Dirac equation is
given by the poles of (\ref{181}). They are given by the poles of
$\Gamma\left(1+\sigma - \frac{Z\alpha\varepsilon}{\lambda}\right)$:
\begin{equation}
  \sqrt{\varkappa^2-Z^2\alpha^2} -
  \frac{Z\alpha\varepsilon}{\sqrt{m^2-\varepsilon^2}}
  = -1, -2, \dots\equiv -n_r,
  \label{19}
\end{equation}
to which must be added, for $\varkappa<0$, the zero of the last term
in the denominator of (\ref{181})\footnote{This term is proportional
  to the sum
  $\left(\sqrt{\varkappa^{2}-\left(Z\alpha\right)^{2}}-\frac{\displaystyle
      Z\alpha\varepsilon}
    {\displaystyle\sqrt{m^{2}-\varepsilon^{2}}}\right)+\left(\varkappa+
    \frac{\displaystyle Z\alpha
      m}{\displaystyle\sqrt{m^{2}-\varepsilon^{2}}}\right)$. It is easy to check
  that, when the first term vanishes, so does the second. Their sum
  increasing monotonically when $\varepsilon$ increases can vanish
  only once.}:
\begin{equation}
  Z\alpha\sqrt{m-\varepsilon}+(\varkappa
  +\sigma)\sqrt{m+\varepsilon}=0
  \;\Rightarrow\;\sqrt{\varkappa^2-Z^2\alpha^2} -
  \frac{Z\alpha\varepsilon}{\sqrt{m^2 -\varepsilon^2}}
  = 0 \equiv n_{r}.
  \label{20}
\end{equation}

The electron bound states at $Z < Z_{\rm cr}$ become therefore
resonances at $Z > Z_{\rm cr}$; the poles of the $S$ matrix
corresponding to the latter describe positron-nucleus scattering. The
trajectory of the ground state energy with growing $Z$ is shown in
Fig.~\ref{fig:e_of_Z} (see also Table~\ref{tab:e_of_Z}).

Let us notice the unusual signs of both real and imaginary parts of
the resonance energy. It was suggested in \cite{Kuleshov} that the
sign of the energy should be reversed, under the claim that the
corresponding state is a resonance in the positron-nucleus
system. Such a sign reversal is usual for holes in the lower continuum
of the Dirac equation: the absence of an electron of energy
$-\varepsilon$ is equivalent to the presence of a positron with energy
$\varepsilon$. Advocating for the same procedure in the case at hands
looks a priori suspicious since the resonances that we found originate
from electron bound energy levels (however, also empty) that dive from
$\varepsilon=+m$ downwards into the lower continuum (and will return
upwards to $+m$ if $Z$ decreases). An interpretation of the phenomenon
in terms of electrons looks therefore more intuitive. In order to
resolve this (apparent) puzzle, we shall solve in the next section the
Dirac equation for positrons, which describes the scattering of a
positron in the upper continuum on a nucleus.

\section{The Dirac equation for positrons: upper continuum wave
  functions and scattering phases in the Coulomb field of a
  supercritical nucleus}
\label{sec:upper}

Changing the sign of $Z\alpha$ in (\ref{eq:potential}), we get instead
of (\ref{eq:3})

\begin{equation}
  \left\{
  \begin{aligned}
    &\frac{d\tilde F}{dr} + \frac{\varkappa}{r} \tilde F - (\varepsilon +m -\tilde V(r)) \tilde G = 0,\\
    &\frac{d\tilde G}{dr} - \frac{\varkappa}{r} \tilde G + (\varepsilon -m -\tilde V(r)) \tilde F = 0,
  \end{aligned}
  \right.
  \label{21}
\end{equation}
where
\begin{subequations}
  \begin{numcases}{\tilde V(r) =}
    \frac{Z\alpha}{r}, & $r > R$,
    \label{eq:pos_potential_r>R}\\
    \frac{Z\alpha}{R}, & $r < R$.
  \end{numcases}
\end{subequations}
Notice that (\ref{eq:3}) gives (\ref{21}) by the set of
transformations $\varkappa\to -\varkappa$,
$\varepsilon\to -\varepsilon$, $F\to\tilde G$ and $G\to\tilde F$.

The states in the upper continuum ($\varepsilon > m$) describe
positron scattering on a nucleus. Since a positron cannot form a bound
state with a positively charged nucleus, one could think that no
resonance at $Z > Z_{\rm cr}$ will occur, nor the resonant behavior of
the scattering phase found in \cite{Kuleshov} and reproduced in
Section~\ref{sec:lower}.

The central issue is therefore to investigate whether bound states and
resonances arise or not in the Dirac equation for positrons
(\ref{21}).

Solving (\ref{21}) at $r < R$ we obtain
\begin{equation}
  \left(
    \begin{array}{l}
      \tilde F \\
      \tilde G
    \end{array}\right)
  = {\rm const}\cdot\sqrt{\tilde\beta r}\cdot
  \left(
    \begin{array}{l}
      \pm J_{\mp(1/2 + \varkappa)} (\tilde\beta r) \\
      J_{\pm(1/2 - \varkappa)} (\tilde\beta r)
      \frac{\tilde\beta}{\varepsilon+m-\frac{Z\alpha}{R}}
    \end{array}
  \right),\; r < R,
  \label{23}
\end{equation}
where $\tilde\beta = \sqrt{(\varepsilon - Z\alpha/R)^2 - m^2}$ and, at
small distances, where the solution (\ref{23}) will be used,
$\tilde\beta \approx \beta \approx Z\alpha/R$. The upper (lower) signs
in (\ref{23}) should be taken for $\varkappa < 0$ ($\varkappa >
0$). Note that the sign of $\tilde F$ is opposite to that of $F$ in
(\ref{5}), while the signs of $\tilde G$ and $G$ coincide.

Substituting in (\ref{21}) the Coulomb potential
(\ref{eq:pos_potential_r>R}) and going to the limit $r \to 0$ we get:
\begin{equation}
  \left.\left(
      \begin{array}{l}
        \tilde F \\
        \tilde G
      \end{array}
    \right)\right|_{r \to 0}
  = \tilde\eta_\sigma r^\sigma \left(
    \begin{array}{c}
      -1 \\
      \frac{-Z\alpha}{\sigma -\varkappa}
    \end{array}
  \right)
  + \tilde\eta_{-\sigma} r^{-\sigma} \left(
    \begin{array}{c}
      -1 \\
      \frac{-Z\alpha}{-\sigma -\varkappa}
    \end{array}
  \right).
  \label{24}
\end{equation}
Note that the sign of $\tilde G$ is opposite to that of $G$ in
(\ref{6}), while the signs of $\tilde F$ and $F$ coincide. Thus, when
matching the ratios of $\tilde F/\tilde G$ from (\ref{23}) and
(\ref{24}) at $r=R$ we obtain equations identical to (\ref{7}),
(\ref{77}) with the change $\eta\to\tilde\eta$.

Like in (\ref{8}), we look for solutions of the form
\begin{equation}
  \left(
    \begin{array}{l}
      \tilde F \\
      \tilde G
    \end{array}
  \right) = \left(
    \begin{array}{c}
       \sqrt{m+\varepsilon} \\
      -\sqrt{m-\varepsilon}
    \end{array}
  \right) {\rm exp}(-\rho/2) \rho^{i\tau}
  \left(
    \begin{array}{c}
      \tilde Q_1 + \tilde Q_2 \\
      \tilde Q_1 - \tilde Q_2
    \end{array}
  \right),
  \label{25}
\end{equation}
where as before $\rho = 2\lambda r = -2ikr$. The expressions for
$\tilde Q_1$ and $\tilde Q_2$ are given by (\ref{eq:9}), where
$Z\alpha$ should be substituted by $-Z\alpha$:
\begin{equation}
  \hspace{-7mm}\left\{
    \begin{aligned}
      \tilde Q_1 &= C\cdot\frac{\frac{iZ\alpha m}{k} + \varkappa}
      {-i\tau-\frac{iZ\alpha\varepsilon}{k}}\cdot{_{1}F_{1}}\left(
        i\tau +
        \frac{iZ \alpha \varepsilon}{k}, 2i\tau + 1, \rho \right) + \\
      &\hspace{40mm} + D\cdot\frac{\frac{iZ\alpha m}{k} +
        \varkappa}{i\tau-\frac{iZ\alpha\varepsilon}{k}}\cdot
      \rho^{-2i\tau}\cdot {_{1}F_{1}}\left(-i\tau +
        \frac{iZ\alpha\varepsilon}{k}, -2i\tau+1,
        \rho\right), \\
      \tilde Q_2 &= C\cdot{_{1}F_{1}}\left(1+i\tau +
        \frac{iZ\alpha\varepsilon}{k}, 2i\tau + 1, \rho \right) +
      D\rho^{-2i\tau}\cdot{_{1}F_{1}}\left( 1-i\tau +
        \frac{iZ\alpha\varepsilon}{k}, -2i\tau+1, \rho \right).
    \end{aligned}
  \right.
  \label{26}
\end{equation}

Since we are interested in resonant states, we demand that only terms
$\propto\exp[ikr]$ (outgoing waves) survive at $r\to\infty$. For
$\varepsilon<m$, $\exp[ikr]$ becomes $\exp[-\lambda r]$, which
describes bound states. In $\tilde Q_1$, the coefficient of the
$\exp[-ikr]$ term, being damped by an extra $1/r$, does not
contribute, and the condition for the terms proportional to
$\exp[-ikr]$ to be absent in $\tilde Q_2$ is
\begin{equation}
  C\cdot\frac{\Gamma(2i\tau)}
  {\Gamma\left(1+i\tau+\frac{iZ\alpha\varepsilon}{k}\right)}
  - D\cdot\frac{\Gamma(-2i\tau)}
  {\Gamma\left(1-i\tau+\frac{iZ\alpha\varepsilon}{k}\right)} = 0.
  \label{27}
\end{equation}

Substituting (\ref{26}) into (\ref{25}) at the limit $r\to 0$, we
reproduce (\ref{24}) for
\begin{equation}
  \begin{aligned}
    \frac{\tilde\eta_\tau}{\tilde\eta_{-\tau}} &
    =  \frac{(-2ik)^{i\tau}}{(-2ik)^{-i\tau}}\cdot
    \frac{\Gamma(-2i\tau)}{\Gamma(2i\tau)}\cdot
    \frac{\Gamma\left(1+i\tau+\frac{iZ\alpha\varepsilon}
        {k}\right)}
    {\Gamma\left(1-i\tau+\frac{iZ\alpha \varepsilon}
        {k}\right)}\cdot
    \frac{Z\alpha\sqrt{m-\varepsilon}
      -(-i\tau+\varkappa)\sqrt{m+\varepsilon}}
    {Z\alpha\sqrt{m-\varepsilon}
      -(i\tau+\varkappa)\sqrt{m+\varepsilon}}.
  \end{aligned}
  \label{28}
\end{equation}

Matching Eqs.~(\ref{28}) and (\ref{77}) yields an equation for the
energies of the resonant states. After the substitution of
($\varkappa, \varepsilon$) by ($-\varkappa, -\varepsilon$), it
coincides with the similar equation that we obtained in
Section~\ref{sec:lower}. Thus, resonances also arise as solutions of
the Dirac equation for positrons, at energies
$\varepsilon=\xi-\frac{i}{2}\gamma,~\xi>m,~\gamma>0$\footnote{One may
  wonder how a positron, being repelled from the positively charged
  nucleus, can form a quasistationary resonance state with it. This
  unusual phenomena is explained in Appendix
  \ref{sec:eff_potential}.}.

After making the same substitutions as in Section~\ref{sec:lower}, we
get equations that coincide with (\ref{171}), (\ref{181})). This
clears the mystery concerning the resonances that we have found
there. Positrons states of negative energies should be interpreted in
terms of electrons. At $Z < Z_{\rm cr}$ we just found
electron--nucleus bound states --- with growing $Z$, the energy of the
bound particle moves from $-m$ (at $Z=0$) to $+m$ (see also footnote
\ref{footnote:pos_bound_states}) and, at $Z > Z_{\rm cr}$ it becomes
complex and located on the second sheet below the right cut.

Equations for the scattering phase $\delta_{\varkappa}$ analogous to
(\ref{eq:phase}),~(\ref{77}),~(\ref{eq:CD}) in Section~\ref{sec:lower}
can be written. They coincide with these equations after changing
$\varkappa\to-\varkappa$ and $\varepsilon\to-\varepsilon$.

It is therefore not necessary to solve the Dirac equation for
positrons as we did in this section. It is enough to note that, after
substitution of $\varepsilon$ by $-\varepsilon$, $\varkappa$ by
$-\varkappa$, $F$ by $\tilde G$ and $G$ by $\tilde F$,
Eq.~(\ref{eq:3}) becomes (\ref{21}) with $V(r)$ converted to
$\tilde V(r)$. In this way, the formulas of Section~\ref{sec:upper}
can be directly deduced from the ones of Section~\ref{sec:lower}.

\section{Conclusions}
\label{sec:conclusions}

In Sections \ref{sec:lower} and \ref{sec:upper}, the scattering of
positrons on a supercritical nucleus was studied. It has the
spectacular resonance behavior discovered in
\cite{MRG1:1972,MRG2:1972,Kuleshov}. In the present paper, results
with an exact dependence on the parameter $m\times R$ have been
obtained on both sheets of the complex energy plane in the form
convenient for numerical evaluation. However, one can hardly hope to
study this phenomenon experimentally: even if a supercritical nucleus
can be produced in heavy ions collisions, its life time will be so
short that one cannot scatter a positron on it, not to mention the
still bigger challenge of making a target with supercritical
nuclei. Let us note that since the elastic scattering matrix was found
to be unitary (the scattering phase is real) there are no inelastic
processes in the positron scattering on supercritical nucleus.

More realistic is the hope to detect the emission of positrons from a
short-lived supercritical nucleus eventually produced in heavy ions
collisions. Indeed we do not agree with the claim made in the abstract
of \cite{Kuleshov} (and in contradiction with \cite{Zeldovich:1971} in
particular) that the spontaneous production of $e^+e^-$ pairs from a
supercritical nucleus does not occur. On the contrary, we believe that
the resonance found in \cite{Kuleshov} in the system
positron---supercritical nucleus is precisely the signal for pair
production. It occurs when, as $Z$ grows, an empty electron level
dives into the lower continuum of the Dirac equation. In the absence
of the nucleus, this empty state in the lower continuum would just
mean the presence of a positron.  The presence of the nucleus makes
the energy of this state complex, and its lifetime is precisely
$1/\gamma$.  In this lapse of time, an electron from the sea with the
same energy $-\xi$ located far from the nucleus can penetrate in its
vicinity. It partially screens the charge of the nucleus and, at the
same time, an empty electron state arises in the Dirac sea. This is
the positron which gets repulsed to infinity by the nucleus.

Let us suppose that solutions of the Dirac equation we get are
approximately valid also when an electron screens nuclear potential,
being embedded in the lower continuum. It means our solutions for the
resonance energy and width are almost valid. It well can be so, since
electric charge of one electron is small and it is situated far from
nucleus, $r \approx 1/m$. So, the obtained width (imaginary part of
energy) is the lifetime of positron in the vicinity of nucleus, which
is already surrounded by diving electron. Therefore this is the
lifetime of the system of nucleus, electron and positron with respect
to positron emission to infinity, so it is an average time of
$e^{+}e^{-}$ pair production (in reality two independent pairs are
produced because of electron spin degeneracy).

The potential barrier which holds the positron in the vicinity of the
nucleus is shown in Fig.~2 of \cite{Zeldovich:1971}; its penetration
time is given by the analytical formulas (4.14, 4.15), and the results
of numerical calculations are shown in Fig.~13 of the same review
paper. We reproduced the curve shown in Fig.~13 from the dependence
$\gamma(Z)$ that we obtained in Section~\ref{sec:lower} for the energy
of the Gamov (quasistationary) state.

Let us finally mention that we agree with the description of the
stable states of a supercritical nucleus made in Section~6 of
\cite{Kuleshov}: empty states in the upper continuum, empty discrete
levels, and occupied states in the lower continuum. The levels of the
lower continuum that get occupied by electrons after the diving
process form the so-called ``charged vacuum''; it has charge $-n$,
where $n$ is the number of these levels. The $n$ positrons that get
emitted compensate for this negative charge. A supercritical nucleus
is no longer naked and its electric charge is partially screened by
these electrons.

Indirect evidence of such a phenomenon is found in graphene physics
\cite{Wang734,NaturePhysics}.

We thank O.V.~Kancheli, V.D.~Mur, V.A.~Novikov, and M.I. Eides for
useful discussions. We are grateful to V.M.~Shabaev who provided us
with references \cite{Maltsev:2014qna,Maltsev2017}. S.G. is supported
by RFBR under grants 16-32-60115 and 16-32-00241, by the Grant of
President of Russian Federation for the leading scientific Schools of
Russian Federation, NSh-9022-2016, and by the ``Dynasty Foundation''.
M.V. is supported by RFBR under grant 16-02-00342. M.V. is grateful to
LPTHE, CNRS and Sorbonne Univesit\'e for hospitality and funding
during the first steps (projet IDEX PACHA OTP-53897) and the last
steps of this work.

\appendix
\section{Functions $Q_1$ and $Q_2$}
\label{sec:Q1Q2}

Substituting (\ref{8}) into the Dirac equations (\ref{eq:3}) we get:
\begin{equation}
  \begin{aligned}
    &\rho(Q_1^\prime + Q_2^\prime) + (i\tau +\varkappa)(Q_1 +Q_2)-\rho Q_2
    + Z\alpha \sqrt{\frac{m-\varepsilon}{m+\varepsilon}} (Q_1 - Q_2)
    = 0,\\
    &\rho(Q_1^\prime - Q_2^\prime) + (i\tau -\varkappa)(Q_1 -
    Q_2)+\rho Q_2 -Z\alpha
    \sqrt{\frac{m+\varepsilon}{m-\varepsilon}} (Q_1 + Q_2) = 0,
  \end{aligned}
\label{A1}
\end{equation}
where a prime means the derivative with respect to $\rho$.

The sum and difference of the two equations (\ref{A1}) give (compare
with Eq.~(36.5) from \cite{BLP}):
\begin{equation}
  \begin{aligned}
    &\rho Q_1^\prime + \left(i\tau -
      \frac{iZ\alpha\varepsilon}{k}\right)Q_1 + \left(\varkappa -
      \frac{iZ\alpha m}{k}\right) Q_2 = 0,\\
    &\rho Q_2^\prime + \left(i\tau - \rho
      +\frac{iZ\alpha\varepsilon}{k}\right)Q_2 + \left(\varkappa +
      \frac{iZ\alpha m}{k}\right) Q_1 = 0.
  \end{aligned}
\label{A2} 
\end{equation}
Eliminating $Q_1$ or $Q_2$ gives
\begin{equation}
  \begin{aligned}
    &\rho Q_1^{\prime\prime} + (2 i\tau +1 -\rho)Q_1^\prime +
    \left(\frac{iZ\alpha \varepsilon}{k} - i\tau\right) Q_1 = 0,\\
    &\rho Q_2^{\prime\prime} + (2 i\tau +1 - \rho)Q_2^\prime +
    \left(\frac{iZ\alpha \varepsilon}{k} - 1 - i\tau\right) Q_2=0.
  \end{aligned}
\label{A3} 
\end{equation}
Unlike in the case of a point-like nucleus, we do not demand here that
the solutions of (\ref{A3}) be regular at $\rho=0$. We accordingly
consider linear superpositions of the two independent solutions of the
second order differential equations (\ref{A3}) with arbitrary
coefficients.

First let us recall that the general solution of the equation
\begin{equation}
  zu^{\prime\prime} + (\gamma - z)u^\prime -\alpha u = 0
  \label{A4}
\end{equation}
is:
\begin{equation}
  u = C_1\cdot{_{1}F_{1}}(\alpha, \gamma, z)
  + C_2\cdot z^{1-\gamma}{_{1}F_{1}}(\alpha -\gamma +1, 2-\gamma, z) ,
  \label{A5}
\end{equation}
where $C_1$ and $C_2$ are arbitrary coefficients while the $_1F_1$ are
the Kummer confluent hypergeometric functions. Thus for the solutions
of (\ref{A3}) we obtain:
\begin{equation}
  \hspace{-3mm}
  \begin{aligned}
    Q_1 & = A\cdot{_{1}F_{1}}\left(i\tau -
      \frac{iZ\alpha\varepsilon}{k}, 2i\tau + 1, \rho \right)+
    B\cdot\rho^{-2i\tau}
    {_{1}F_{1}}\left( -i\tau -\frac{iZ\alpha\varepsilon}{k}, -2i\tau+1, \rho \right),\\
    Q_2 & = C\cdot{_{1}F_{1}}\left(1 +i\tau - \frac{iZ\alpha
        \varepsilon}{k}, 2i\tau + 1, \rho \right) +
    D\cdot\rho^{-2i\tau}{_{1}F_{1}}\left(1 -i\tau
      -\frac{iZ\alpha\varepsilon}{k}, -2i\tau+1, \rho \right),
  \end{aligned}
\label{A6}
\end{equation}
where $A$, $B$, $C$, and $D$ are arbitrary coefficients.

At small $z$, $_1F_1=1+\mathcal{O}(z)$. Substituting the expansions of
(\ref{A6}) at small $\rho$ into the first equation in (\ref{A2})
determines $A$ and $B$, respectively, in terms of $C$ and $D$:
\begin{align}
  &\left( i\tau - \frac{iZ\alpha\varepsilon}{k}\right) A
    + \left(\varkappa - \frac{iZ\alpha m}{k}\right) C = 0,
    \label{A7}\\
  &\left( -i\tau - \frac{iZ\alpha\varepsilon}{k}\right) B
    + \left(\varkappa - \frac{iZ\alpha m}{k}\right) D = 0.
\label{A8}
\end{align}
Plugging then $A$ and $B$ obtained from (\ref{A7}) and (\ref{A8}) into
(\ref{A6}) yields (\ref{eq:9}).

\section{The scattering phase according to \cite{Kuleshov}}
\label{sec:real_phase}

Considering real $\varepsilon<-m$ below the left cut we can rewrite
the expression for the scattering phase in a more compact form.

Let us introduce the following notations equivalent to those used in
\cite{Kuleshov}:
\begin{align}
  \label{eq:kuleshov_notations}
  \exp(i\varphi) &=
  e^{2i\theta}\frac{(2k)^{-i\tau}\Gamma(2i\tau)}
  {(2k)^{i\tau}\Gamma(-2i\tau)},\\
  a &=
  \frac{Z\alpha\sqrt{m-\varepsilon}
    +(-i\tau+\varkappa)\sqrt{m+\varepsilon}}
  {\Gamma(1-i\tau - \frac{iZ\alpha\varepsilon}{k})},\\
  b &=
  \frac{Z\alpha\sqrt{m-\varepsilon}
    -(-i\tau+\varkappa)\sqrt{m+\varepsilon}}
  {\Gamma(1-i\tau + \frac{iZ\alpha\varepsilon}{k})}.
\end{align}

With these notations the approximate ratio $C/D$ defined by
(\ref{eq:CD}) can be written in the following way:
\begin{align}
  \label{eq:CD_Kuleshov_numerator}
  C/D
  &=e^{i\varphi-\pi\tau}\cdot\frac{a^{*}}{b}\cdot
    \frac{\Gamma\left(-2i\tau\right)}
    {\Gamma\left(2i\tau\right)}\cdot
    \frac{\Gamma\left(i\tau+\frac{iZ\alpha\varepsilon}{k}\right)}
    {\Gamma\left(-i\tau+\frac{iZ\alpha\varepsilon}{k}\right)}=\\
  \label{eq:CD_Kuleshov_denominator}
  &=e^{i\varphi-\pi\tau}\cdot\frac{b^{*}}{a}\cdot
    \frac{\Gamma\left(-2i\tau\right)}
    {\Gamma\left(2i\tau\right)}\cdot
    \frac{\Gamma\left(1+i\tau-\frac{iZ\alpha\varepsilon}{k}\right)}
    {\Gamma\left(1-i\tau-\frac{iZ\alpha\varepsilon}{k}\right)},
\end{align}
where we used
\begin{equation}
  \frac{\left(-i\right)^{-i\tau}}{\left(-i\right)^{i\tau}}=e^{-\pi\tau},
\end{equation}
and
\begin{align}
  \left(-i\tau-\frac{iZ\alpha\varepsilon}{k}\right)
  &
    \left(Z\alpha\sqrt{m-\varepsilon}+
    \left(i\tau-\varkappa\right)\sqrt{m+\varepsilon}\right)=\nonumber\\
  \nonumber
  &=-i\tau\left(Z\alpha\sqrt{m-\varepsilon}
    +\left(i\tau-\varkappa\right)\sqrt{m+\varepsilon}\right)
    -iZ\alpha\varepsilon\left(\frac{Z\alpha}{i\sqrt{m+\varepsilon}}
    +\frac{i\tau-\varkappa}{i\sqrt{m-\varepsilon}}\right)=\\ \nonumber
  &=\frac{-i\tau\left(i\tau-\varkappa\right)\left(m+\varepsilon\right)
    -\left(Z\alpha\right)^{2}\varepsilon}{\sqrt{m+\varepsilon}}
    +\frac{-i\tau\left(Z\alpha\right)\left(m-\varepsilon\right)
    -Z\alpha\varepsilon\left(i\tau-\varkappa\right)}{\sqrt{m-\varepsilon}}=\\ \nonumber
  &=\frac{\left(Z\alpha\right)^{2}m
    +\varkappa\left(i\tau-\varkappa\right)\left(m+\varepsilon\right)}
    {\sqrt{m+\varepsilon}}
    +\frac{-\left(i\tau-\varkappa\right)Z\alpha m-\varkappa
    Z\alpha\left(m-\varepsilon\right)}{\sqrt{m-\varepsilon}}=
    \nonumber \\
  &=Z\alpha
    m\left(\frac{Z\alpha}{\sqrt{m+\varepsilon}}+
    \frac{-i\tau+\varkappa}{\sqrt{m-\varepsilon}}\right)
    -\varkappa\left(Z\alpha\sqrt{m-\varepsilon}
    +\left(-i\tau+\varkappa\right)\sqrt{m+\varepsilon}\right)=\nonumber\\ 
  &=\left(Z\alpha\sqrt{m-\varepsilon}+
    \left(-i\tau+\varkappa\right)\sqrt{m+\varepsilon}\right)\left(\frac{iZ\alpha
    m}{k}-\varkappa\right),
\end{align}
where we used the relation
\begin{align}
  \left(-i\tau-\varkappa\right)\left(i\tau-\varkappa\right)=
  \tau^{2}+\varkappa^{2}=\left(Z\alpha\right)^{2}.
\end{align}

Then, the scattering phase $\delta_{\varkappa}$ can be written as
follows (by substituting (\ref{eq:CD_Kuleshov_numerator}) and
(\ref{eq:CD_Kuleshov_denominator}) into the numerator and the
denominator of (\ref{eq:phase}) respectively):

\begin{equation}
  e^{2i\delta_{\varkappa}} =
  -\frac{{\rm exp}\left(\frac{\pi\tau}{2}+\frac{i\varphi}{2}\right)a^*
    -{\rm exp}\left(-\frac{\pi\tau}{2}-\frac{i\varphi}{2}\right)b}
  {{\rm exp}\left(\frac{\pi\tau}{2}-\frac{i\varphi}{2}\right)a
    -{\rm exp}\left(-\frac{\pi\tau}{2} +\frac{i\varphi}{2}\right)b^*}.
  \label{13}
\end{equation}

In eq.(22) of \cite{Kuleshov} the phase $\delta_{\varkappa}$ is
expressed through the ratio $f^*/f$, where $f$ is the Jost
function. Our result differs from eq.(23) of \cite{Kuleshov} by the
substitution $\varphi\to\varphi/2$ (it seems that there is a misprint
in \cite{Kuleshov}).

\section{Qualitative explanation of the resonance phenomena in the
  $e^+ N^+$ system}
\label{sec:eff_potential}

The effective potential for an electron in the field of a
supercritical nucleus is derived in \cite{Zeldovich:1971} from the
Dirac equation, for $\varepsilon\approx-m$. It is attractive at short
distances, repulsive at large distances, with a Coulomb barrier in
between. We derive below, in a similar way, the effective potential
for a positron in the field of a similar nucleus, in the vicinity of
$\varepsilon=+m$.

As already noticed at the beginning of Section~\ref{sec:upper}, the
Dirac equation (\ref{eq:3}) for electrons becomes (\ref{21}) for
positrons after the following substitutions:
\begin{equation}
\varkappa\to-\varkappa, \;\; \varepsilon\to-\varepsilon,
\;\; F \to \tilde G,
\;\; G \to \tilde F,
\;\; V(r) \to \tilde V(r) = -V(r).
\label{C1}
\end{equation}
To proceed like in \cite{Zeldovich:1971}, we deduce the second order
differential equation satisfied by $\tilde G$ from (\ref{21}), which
is
\begin{equation}
  \label{eq:G_shred}
  \tilde G''
  +\frac{\tilde V'}{\varepsilon-m-\tilde V}\left(\tilde
    G'-\frac{\varkappa}{r}\tilde G\right)
  +\left(\left(\varepsilon-\tilde V\right)^{2}-m^{2}
    +\frac{\varkappa\left(1-\varkappa\right)}{r^{2}}\right)\tilde G=0,
\end{equation}
in which ``$~'~$'' means here derivation with respect to $r$. In order
to transform this equation into a Schr\"{o}dinger-like equation, the
following change of variables must be operated
\begin{equation}
  \label{eq:G_to_chi}
  \tilde G=\chi\sqrt{m-\varepsilon+\tilde V}.
\end{equation}

Thus, we get:
\begin{equation}
  \label{eq:chi_shred}
  \chi''+ k^{2}\chi=0,
\end{equation}
where $k^{2}=2m\left(E-U\right)$,
$E=\frac{\varepsilon^{2}-m^{2}}{2m}$. The effective potential is seen
to be made of two terms: $U=U_{1}+U_{2}$, where:
\begin{equation}
  \label{eq:U1}
  U_{1}=\frac{\varepsilon}{m}\tilde V-\frac{1}{2m}\tilde V^{2}
  -\frac{\varkappa\left(1-\varkappa\right)}{2mr^{2}},
\end{equation}
and
\begin{equation}
  \label{eq:U2}
  U_{2}=\frac{\tilde V''}{4m\left(\varepsilon-m-\tilde V\right)}
  +\frac{3}{8m}\frac{\left(\tilde
      V'\right)^{2}}{\left(\varepsilon-m-\tilde V\right)^{2}}
  +\frac{\varkappa\tilde V'}{2mr\left(\varepsilon-m-\tilde V\right)}.
\end{equation}
It coincides with the equation obtained in \cite{Zeldovich:1971} after
the substitution $\varepsilon\to-\varepsilon$,
$\varkappa\to-\varkappa$ and $\tilde V\to-V$.

We are interested in positrons with $\varepsilon\approx m$. At large
distances the first term in $U_1$ dominates, and describes the
repulsion of the positron by the nucleus. For the ground state
$\varkappa=1$ the centrifugal term in $U_1$ vanishes. Finally, for
$\varepsilon=m$ and $\varkappa=1$, we get from (\ref{eq:U1}) and
(\ref{eq:U2})
\begin{equation}
  \label{eq:eff_potential}
  U = \frac{Z\alpha}{r} + \frac{3-4(Z\alpha)^2}{8mr^{2}}.
\end{equation}
At short distances the terms $\propto 1/r^2$ dominates and, for a
supercritical nucleus, they lead to attraction, while the Coulomb term
dominates at $r\geq1/m$. This attractive force explains the existence
of resonances in the $e^{+}N^{+}$ system while bound state cannot
exist due to the narrowness of the well.

Let us note that the fall to the center occurs only for $Z\alpha>1$
when the coefficient in front of the term $\propto -1/r^2$ becomes
larger than $1/8m$ (see \cite{LL3}, eq. (35.10)). We are grateful to
V.A. Novikov who brought our attention to this feature. In the problem
under consideration the finite nucleus size prevents the fall to the
center.

\bibliographystyle{apsrev4-1}
\bibliography{references}

\end{document}